\documentclass[journal,twoside]{IEEEtran}
\usepackage{cite}
\usepackage{mathptm}    % to use math fonts
\usepackage{mathptmx}  
\usepackage{stfloats}
\usepackage{multirow}
\usepackage{amssymb}
\usepackage{pslatex}	
\usepackage[cmex10]{amsmath}
\usepackage[dvips]{graphicx}
\usepackage[normalem]{ulem}
\begin{document}

\title{Circulant Matrix Representation of PN-sequences with Ideal Autocorrelation Property}

\author{Mohammad~J.~Khojasteh, Morteza~H.~Shoreh, and Jawad~A.~Salehi, \emph{Fellow, IEEE}
\thanks{Part of this paper was supported by  Iran National Science Foundation (INSF).}
\IEEEauthorblockA{\\Optical Networks Research Laboratory\\Department of Electrical Engineering \\Sharif University of Technology,Tehran, Iran\\
Email: jasalehi@sharif.edu}}

\maketitle

\begin{abstract}
   In this paper, we investigate PN-sequences with ideal autocorrelation property and the consequences of this property on the number of $+1$s and $-1$s and run structure of sequences. We begin by discussing and surveying about the length of  PN-sequences with ideal autocorrelation property. From our discussion and survey we introduce circulant matrix representation of PN-sequence. Through   circulant matrix representation we obtain system of non-linear equations that lead to ideal autocorrelation property. Rewriting PN-sequence and its autocorrelation property in $\{0, 1\}$  leads to a definition based on Hamming weight and Hamming distance and hence we can easily prove some results on the PN-sequences with ideal autocorrelation property.
  
\end{abstract}


%\newpage
%\setlength{\baselineskip}{1.8\baselineskip}

\section{Introduction}
\IEEEPARstart{P}{seudo} noise sequences (PN-sequences) are codes that are considered to have correlation and spectrum properties similar to random sequences, although they are deterministically generated. There are many versions of PN-sequences with different definitions, approaches and applications  such as, maximal-length sequences (m-sequences) \cite{Golomb}, Gold codes \cite{Gold}, zero correlation zone sequences (ZCZ)\cite{zcz}, etc.
In general, m-sequences are among the most important PN-sequences since they satisfy randomness postulates stated by Golomb \cite{Algebra}, namely, \textit{ideal autocorrelation property}, \textit{balance property}, and \textit{run property}. In further work by Golomb he makes the following conjecture \cite{Algebra}, which is still considered open: ``\textit{The only binary sequences satisfying the three randomness postulates are m-sequences.}"\cite{Algebra}. 

The correlation between all non-zero cyclic shifts of an m-sequence is almost zero (ideal autocorrelation property) \cite{Magazine}, so they can be used as sequences with excellent autocorrelation function. Sequences with ideal autocorrelation property are in one-to-one correspondence with Paley-Hadamard difference 
sets \cite{Glomb2}. A general algorithm for constructing these classes of sequences for any arbitrary length $n$ is not known so far.

Golomb states another conjecture on the existence of  Paley-Hadamard difference sets that is if $n$, the length of Paley-Hadamard difference sets, is equal to $4k+3$, then it should be either a prime number, or $n$ must be the product of twin primes or it should be in the form of $2^k-1$, where $k$ is a positive integer \cite{conjecture2}. 
To the best of our knowledge when $n$ is a prime number only Legendre sequences \cite{Legendre} and sextic residue construction \cite{Hall} are known. The other known sequences with ideal autocorrelation property are; Jacobi symbol\cite{Jacob} for $n=p(p+2)$, and m-sequences\cite{Golomb}, Gordon-Mills-Welch (GMW)  sequences \cite{GMW}  and  miscellaneous instances\cite{conjecture} for $n=2^k-1$.

Golomb believes that the existence of miscellaneous examples gives a clue for further investigating the truth of his conjecture about Paley-Hadamard difference sets. Three of these examples were founded in 1967 for $n=127$, and a few years later two and three  examples   were found for $n=255$, and $n=511$, respectively. In 1998 in \cite{Binary}, the authors constructed five new classes of binary sequences with ideal autocorrelation by exhaustive search 
for $n=2^k-1$ for all $k\leq10$, and proposed a few more conjectures on the general construction of these sequences and their corresponding difference sets.
  
In many applications generalizing the length of PN-sequences is critical  such as in spectrum fragmented cognitive radio networks [14,15], where the sequences should have a wide range of lengths because of the number of available sub-carriers differ in various conditions. Hence  in many advanced communications systems, codes with various lengths are needed.

In generalizing the length of PN-sequences we begin by proposing the circulant matrix representation of PN-sequences. The idea of using circulant matrix representation to construct a desired sequence  was first used by Alem and Salehi in \cite{Alem} in order to represent Optical Orthogonal Code (OOC). In \cite{Alem} the search space  is spectrally classified  using circulant matrix representation of OOCs, followed by a  group action that introduces an efficient partitioning algorithm.

The rest of this paper is organized as follows; in Section II, circulant matrix representation of PN-sequences is proposed. In Section III, based on circulant matrices representation, a system of $n$ non-linear equations is proposed that can be used to justify ideal autocorrelation property of PN-sequences.  Then a new perspective arises by transferring circulant matrix of PN-sequences to $\{0,1\}$ domain  which leads to a better understanding of these sequences discussed in Section IV.  The run structure of the desired sequences are investigated in Section V. Finally, Section VI, summarizes the results and concludes the paper. 

\section{Circulant Matrix Representation of PN-sequences}
Lets denote a PN-sequence via a codeword, $x=(x_0,x_1,\ldots,x_{n-1})$. In most technical literature a codeword $x$ is said to have ideal autocorrelation property if it has the following autocorrelation function [4,13]
 \begin{align}
R_x(\tau) = \left\{ \begin{array}{rcl}
n & \mbox{for}~
 \tau \equiv 0 \bmod {n}	  \\ -1 &  otherwise 
\end{array}\right.
\end{align}
where $R_x(\tau)$ is defined as 
\begin{align}
R_x(\tau)=\sum\limits_{l=0}^{n-1} x_{l}x_{l\oplus\tau}.
\end{align}
and $\oplus$ is n-module addition.

Herein, we recognize that in bipolar codewords, $\pm1$s  average out each other in order to construct an impulse shape autocorrelation function \cite{OOC}. In general PN-sequences with ideal autocorrelation property are similar to OOCs, since  both have cyclic structure with cyclic ideal autocorrelation property. The idea of using outer product matrix to design a new searching algorithm to obtain OOC codewords was first proposed in \cite{Charm} by Charmchi and Salehi, where the authors attempt, successfully, remove the bottleneck of designing and generating OOCs with  certain code lengths.
In \cite{Alem}, in order to develop search algorithm in designing OOCs the authors do an in depth search for finding  appropriate types of matrices to representing the characteristics of OOCs. In the following definitions, the circulant matrix representation of PN-sequence is introduced, as in \cite{Alem}, whereby displaying all possible cyclic shifts of a codeword in a circulant matrix. 

\textit{Definition~1:} The circulant matrix representation of every codeword $x=(x_0,x_1,\ldots,x_{n-1})$ as a binary PN-sequence~($x_l\in \{ \pm 1\}$ for ~$0\leq l \leq n-1$)~is defined as follows

\begin{align}
 A_x=A_{(x_0,x_1,\ldots,x_{n-1})}=\left[ \begin{array}{cccccc}
  x_0  & x_1  & \ldots  & x_{n-1}   \\ x_{n-1} & x_0  & \ldots & x_{n-2} \\ \vdots & \vdots & \ldots & \vdots \\ x_1 & x_2 & \ldots & x_0   \end{array} \right]. 
\end{align}
%, thus $A_{x_{i,j}}$ is given by \cite{Circulant}
%\begin{align}
%A_{x_{i,j}}=x_{(j-i)~ mod ~n} 
%\end{align}

Every row of a circulant matrix is a cyclic shift of it's above row \cite{Davis}. From (1), (2) and (3) it becomes evident that the condition of ideal autocorrelation for $x=(x_0,x_1,\ldots,x_{n-1})$ and its circulant matrix $A_x$ is presented as follows;
\begin{align}
A_xA_x^T=nI_n+E_n=A_{(n,-1,\ldots,-1)} \label{5}
\end{align} 
where $I_n$ represents the identity matrix of order $n$ and if $J_n$ denotes an $n \times n$ all-ones matrix (every element of $J_n$ is equal to $1$) then 
\begin{align}
E_n=-J_n+I_n. \label{6}
\end{align}
\textit{Example~1:}
If $x=(-1,-1,+1)$ (m-sequence of length 3) then
\begin{align}
A_x=\left[ \begin{array}{ccc}
  -1  & -1  &   +1    \\  +1 & -1  & -1 \\ -1 &   +1 & -1    \end{array} \right] 
\end{align}
and
\begin{align}
A_xA_x^T=\left[ \begin{array}{ccc}
  +3  & -1  &   -1    \\  -1 & +3  & -1 \\ -1 &   -1 & +3    \end{array} \right]=3I_3+(-J_3+I_3) 
\end{align}
\section{Properties of Circulant Matrices and the corresponding Non-Linear System of Equations}
The properties of circulant matrices are well known and easily derived in \cite{Circulant}. The matrix in (3) has eigenvectors, and    eigenvalues that are as follows;
\begin{align}
v_{m}=\frac{1}{\sqrt{n}}(1,e^{\frac{-j 2\pi  m}{n}},\ldots,e^{\frac{-j 2\pi  m(n-1)}{n}})^T
\end{align}
\begin{align}
\lambda_{x_m} =\sum\limits_{l=0}^{n-1} x_l e^{-j2\pi\frac{ ml}{n}}
\end{align}
where, $m =0,1,\ldots, n-1$.

 If $U_n$ is an $n \times n$ matrix that has the eigenvectors as columns placed in order (Fourier unitary matrix) and $\Psi=diag(\lambda_{x_m})$ then $A_x=U_n \Psi_{x} U_n^*$. Also matrices that have this eigenvector matrix are circulant \cite{Crespo}.

%\textit{ Example~2:}
%$U_3$ for all the circulant matrices of order 3 can be written as follows:
%\begin{align}
%%U_3=\frac{1}{\sqrt{3}} \left[ \begin{array}{ccccc}
 % 1  & 1    & 1   \\ 1 & e^{-j\frac{2\pi }{3}}  & e^{-j\frac{4\pi }{3}} \\ 1 & e^{-j\frac{4\pi }{3}} &  e^{-j\frac{8\pi }{3}}   \end{array} \right] 
%\end{align}

In order to proceed further we need one more property about circulant matrix. If $x=(x_0,x_1,\ldots,x_{n-1})$ and $y=(y_0,y_1,\ldots,y_{n-1})$ then
\begin{align}
A_xA_y=A_yA_x=U_n \Psi U_n^* \label{yek}
\end{align}
where, $\Psi = diag(\lambda_{x_m} \lambda_{y_m})$ and $A_xA_y$ is also circulant matrix.
If $y=(x_0,x_{n-1},\ldots,x_{1})$, then
\begin{align}
A_xA_x^T=A_xA_y
\end{align}
So by (\ref{yek})
 \begin{align}
 A_xA_x^T=U_n \Psi U_n^* \label{14}
 \end{align}
and $\lambda_{x_m} \times \lambda_{y_m} $ calculated in (\ref{17}). From (\ref{5}) and (\ref{14}) we have

\begin{align}
U_n \Psi U_n^*=nI_n+E_n \nonumber \\
\end{align}
Since $U_n$ is unitary matrix ($UU^*=I$) so
\begin{align}
\Psi=U_n^*(nI_n+E_n)U_n &=nI_n+U_n^*E_nU_n \nonumber \\
\Psi- nI_n&=U_n^*E_nU_n \label{16}
\end{align}
$\Psi- nI_n$ (the left hand side of (\ref{16})) is obtained as in (\ref{18}), and (\ref{19}). 

\begin{table*}
\begin{align}
\lambda_{x_m} \lambda_{y_m}=(x_0+x_1 e^{-j\frac{2 \pi m}{n}}+x_2 e^{-j\frac{2 \pi  m(2)}{n}}&+\cdots  +x_{n-1} e^{-j\frac{2 \pi  m(n-1)}{n}})(x_0+x_{n-1} e^{-j\frac{2 \pi  m}{n}}+x_{n-2} e^{-j\frac{2 \pi  m(2)}{n}}+\cdots+x_1 e^{-j\frac{2 \pi  m(n-1)}{n}}) \nonumber\\ \label{17}
&=x_0^2+x_1^2+\cdots+x_{n-1}^2+2\sum\limits_{l>r} x_lx_r cos(\frac{2\pi m }{n}(l-r)) 
\end{align}

\begin{align} \label{18}
 \Psi - nI_n=\left[ \begin{array}{cccccc} 
  \sum\limits_{l=0}^{n-1} x_l^2 +2\sum\limits_{l>r} x_l x_r -n & 0  & \ldots  & 0   \\ 0 & \sum\limits_{l=0}^{n-1} x_l^2 +2\sum\limits_{l>r} x_l x_r cos(\frac{2 \pi }{n}(l-r))-n & \ldots & 0 \\ \vdots & \vdots & \vdots & \vdots \\ 0 & 0 & \ldots & \sum\limits_{l=0}^{n-1} x_l^2 +2\sum\limits_{l>r} x_l x_r cos(\frac{2 \pi (n-1)}{n}(l-r))-n   \end{array} \right] 
\end{align}
\begin{align} \label{19}
=\left[ \begin{array}{cccccc}
   2\sum\limits_{l>r} x_l x_r  & 0  & \ldots  & 0   \\ 0 & 2\sum\limits_{l>r} x_l x_r cos(\frac{2 \pi }{n}(l-r)) & \ldots & 0 \\ \vdots & \vdots & \vdots & \vdots \\ 0 & 0 & \ldots & 2\sum\limits_{l>r} x_l x_r cos(\frac{2 \pi (n-1)}{n}(l-r))   \end{array} \right]
\end{align}
\end{table*}
There is  a fact about  orthogonality of the complex exponentials\cite{Circulant}
\begin{align}
\sum\limits_{m=0}^{n-1} e^{j\frac{2\pi  m l} {n}}
 = \left\{ \begin{array}{rcl}
n 
& l~mod~n = 0 \\ 0  & otherwise 
\end{array}\right. .
\end{align}

So if $n$ is a prime number then we can easily rewrite the right hand side of (\ref{16}) by substituting $E_n$ from (\ref{6}) 
\begin{align}
U_n^*E_nU_n=U_n^*(-J_n+I_n)U_n=I_n-U_n^*J_nU_n
\end{align}
thus,
\begin{align}
\Psi - nI_n=\left[ \begin{array}{cccccc}
   1-n  & 0  & \ldots  & 0   \\ 0 & 1 & \ldots & 0 \\ \vdots & \vdots & \vdots & \vdots \\ 0 & 0 & \ldots & 1   \end{array} \right] 
\end{align}
which leads to the following system of non-linear  equations
\begin{align}
  \left\{ \begin{array}{rcl}
 \sum\limits_{l>r} x_l x_r=& \frac{1-n}{2} \\  \sum\limits_{l>r} x_l x_r cos(\frac{2 \pi }{n}(l-r))=& 0.5  \\ \vdots \\ \sum\limits_{l>r} x_l x_r cos(\frac{2 \pi (n-1)}{n}(l-r))=& 0.5
\end{array}\right.
\end{align}

Considering the properties of cosine function, these non-linear equations  are dependent, hence, there is no need to solve more than $(n+1)/2$ equations as follows

\begin{align}
  \left\{ \begin{array}{rcl} \label{24}
 \sum\limits_{l>r} x_l x_r=& \frac{1-n}{2} \\  \sum\limits_{l>r} x_l x_r cos(\frac{2 \pi }{n}(l-r))=& 0.5  \\ \vdots \\ \sum\limits_{l>r} x_l x_r cos(\frac{\pi (n-1)}{n}(l-r))=& 0.5
\end{array}\right.
\end{align}
Considering the following equation
 \begin{align}
  (\sum\limits_{l=0}^{n-1} x_l)^2=2\sum\limits_{l>r} x_l x_r +\sum\limits_{l=0}^{n-1} x_l^2 =1
 \end{align}
  the first equation in (\ref{24}) is equivalent to $\sum\limits_{l=0}^{n-1} x_l=\pm 1$.

  \textit{ Corollary:} 
  The ideal autocorrelation property leads to balance property.
   
 In order to find sequences with ideal autocorrelation property, we need to search balanced $\{\pm 1\}^n$ and find codewords satisfying equations in (\ref{24}).

\textit{Example~2}:
As an example for $n=7$ equations in (\ref{24}) for $d_i$ where $i=1,\ldots,6$ are as follows
\begin{align}
 d_1&=x_6x_5+x_5x_4+x_4x_3+x_3x_2+x_2x_1+x_1x_0\nonumber\\
 d_2&=x_6x_4+x_5x_3+x_4x_2+x_3x_1+x_2x_0\nonumber\\
 d_3&=x_6x_3+x_5x_2+x_4x_1+x_3x_0\nonumber\\
 d_4&=x_6x_2+x_5x_1+x_4x_0\nonumber\\
 d_5&=x_6x_1+x_5x_0\nonumber\\
 d_6&=x_6x_0
\end{align}
 reduce to;
\begin{align}
\left[ \begin{array}{cccccc} \label{27}
   cos(\frac{2 \pi}{7})  & \ldots  & cos(6 \times \frac{2 \pi}{7})   \\ \vdots & \vdots & \vdots  \\ cos(6 \times \frac{2 \pi}{7})  & \ldots & cos(36 \times \frac{2 \pi}{7}) \\   \end{array} \right] \left[ \begin{array}{c}
         d_1   \\ \vdots  \\ d_6 \\   \end{array} \right]=\left[ \begin{array}{c}
                  0.5  \\ \vdots  \\ 0.5 \\   \end{array} \right]
\end{align}

Due to the property of cosine function, the first, second, and third columns and rows of above $6 \times 6$ matrix are respectively equal to sixth, fifth and fourth columns and rows. Therefore, (\ref{27}) is rewritten as follows
\begin{align}
\left[ \begin{array}{cccccc} \label{28}
   cos(\frac{2 \pi}{7})  & cos( \frac{4 \pi}{7})  & cos( \frac{6 \pi}{7})   \\ cos(\frac{4 \pi}{7}) & cos(\frac{8 \pi}{7}) & cos(\frac{12 \pi}{7})  \\ cos( \frac{6 \pi}{7})  & cos(\frac{12 \pi}{7}) & cos( \frac{18 \pi}{7}) \\   \end{array} \right] \left[ \begin{array}{c}
         d_1+d_6   \\ d_2+d_5  \\ d_3+d_4 \\   \end{array} \right]=\left[ \begin{array}{c}
                  0.5  \\ 0.5  \\ 0.5 \\   \end{array} \right] 
\end{align}
Multiplying the inverse of the $3 \times 3$ matrix on the left of (\ref{28}); we obtain the following expressions;
\begin{align}\label{29}
d_1+d_6&=x_6x_5+x_5x_4+x_4x_3+x_3x_2+x_2x_1+x_1x_0+x_6x_0=-1\\    \nonumber
d_2+d_5&=x_6x_4+x_5x_3+x_4x_2+x_3x_1+x_2x_0+x_6x_1+x_5x_0 =-1\\ \nonumber
d_3+d_4&=x_6x_3+x_5x_2+x_4x_1+x_3x_0+x_6x_2+x_5x_1+x_4x_0=-1  \nonumber
\end{align}

Solving for (\ref{29}), (\ref{30}) and (\ref{31}) in balance $n$-tuples is sufficient for finding sequences with ideal autocorrelation property. On the other hand, this equations are the multiplication of codeword with first, second and third circular shift, respectively.

\textit{Corollary:} As expected the sequences with ideal autocorrelation property are  solutions to the following non-linear equation system in balanced $n$-tuples of $\{{\pm 1}\}$
\begin{align}
  \left\{ \begin{array}{rcl}
 \sum\limits_{l=0}^{n-1} x_{l}x_{l\oplus 1}=-1   \\ \vdots \\
  \sum\limits_{l=0}^{n-1} x_{l}x_{l\oplus \frac{n-1}{2}}=-1
\end{array}\right.
\end{align} 

Examples of PN-sequences with ideal autocorrelation property can be find in Table I.
\section{Transformation to Domain of $\{0,1\}$}
In this section, we investigate PN-sequences by transferring the $\{\pm1\}$ to $\{0,1\}$, and then discuss the corresponding consequences. If we define the following mapping; 
\begin{align}
\theta:  \{-1,1\}^n &\rightarrow \{0,1\}^n\nonumber\\
(x'_0,\ldots,x'_i,\ldots,x'_{n-1})&=\theta(x_0,\ldots,x_i,\ldots,x_{n-1})\nonumber\\
 =(\frac{1-x_0}{2},\ldots,& \frac{1-x_i}{2},\ldots,\frac{1-x_n}{2}) 
\end{align}
Then the autocorrelation function of $x$ can be written as follows [13,22];
\begin{align}
R_x(\tau)=\sum\limits_{l=0}^{n-1} x_{l}x_{l\oplus\tau}
=\sum\limits_{l=0}^{n-1} (-1)^{x'_{l}+x'_{l\oplus\tau}}\\
=n-2 \omega(x' \oplus T^\tau(x'))  \nonumber
\end{align}
where $\omega(x')$ denotes the Hamming weight of $x'$, and $T^\tau$ represents $\tau$ cyclic shift to the left.
Hence
 \begin{align}
\omega(x' \oplus T^\tau(x')) = \left\{ \begin{array}{rcl}
0 & \mbox{for}~
 \tau \equiv 0 \bmod {n}	  \\ \frac{n+1}{2} &  otherwise 
\end{array}\right .
\end{align}
thus, every two different rows of $A_{x'}$ in $\frac{n+1}{2}$ columns have different value and in $\frac{n-1}{2}$ columns have the same value.
If $x=(x_0,x_1,\ldots,x_{n-1})$ satisfies ideal autocorrelation property, then the sequence $y=(y_0,y_1,\ldots,y_{n-1})=(-x_0,-x_1,\ldots,-x_{n-1})$ also satisfies this property. So without loss of generality suppose $\sum\limits_{i=0}^{n-1} x_i=-1$.  Hence in the columns of every two different rows of $A_{x'}$,  the $(1,1)$ pairs  appears once  more than $(0,0)$ pairs. Eventually there are $\frac{n-3}{4}$ pairs of $(0,0)$ in columns of every two different rows of $A_{x'}$.
 
\textit{ Example~3:}
If $x=(-1,-1,-1,1,-1,1,1)$, then $x'=(1,1,1,0,1,0,0)$ and $Tx'=(0,1,1,1,0,1,0)$ have four pairs of $(1,0)$, two pairs of $(1,1)$ and one pair of $(0,0)$ in their columns.
 
 From the above discussion the following results can be obtained.
 
\textit{Corollary~1:} There is no sequences with ideal autocorrelation property of the length  $2k$ or $4k+1$.

\textit{Corollary~2:} The ideal autocorrelation property is given by
 \begin{align}
 A_{x'}A_{x'}^T&=\frac{n+1}{2}I_n+ \frac{n+1}{4}(J_n-I_n) \nonumber \\
 &=A_{(\frac{n+1}{2},\frac{n+1}{4},\ldots,\frac{n+1}{4})}
 \end{align}
 
%\textit{Example~4:}
%If $x=(-1,-1,+1)$, then
%\begin{align}
% A_{x'}A_{x'}^T&=\left[ \begin{array}{ccc}
%   1  & 1  &   0    \\  0 & 1  & 1 \\ 1 &   0 & 1    \end{array} \right] \left[ \begin{array}{ccc}
  % % %   1 & 0  &  1    \\  1 & 1  & 0 \\ 0 & 1 & 1    \end{array} \right] \\ \nonumber
  %   &=2I_n+1(J_n-I_n)=A_{(2,1,1)}
% \end{align}
 \textit{ corollary~3:} 
 A PN-sequence of length $n$ with ideal Autocorrelation can be seen as a family of codewords, $\{0,1\}^n$,  weighting $\frac{n+1}{2}$ that are cyclic shifts of each other with Hamming distances  equals to $\frac{n+1}{2}$ amongst each other.
\section{Run Structure}
Consider the codeword $x=(x_0,x_1,\ldots,x_{n-1})$, a run of length  $f$ is a block of consecutive $1$s or $-1$s in codeword that is not contained in a larger block of $1$s or $-1$s, and is denoted by $R_f$. Furthermore, let $N(R_f)$ to denote the number of the runs of length $f$. The codeword $x$ has the run property \cite{Algebra}, if 
\begin{align} \label{38}
 \lfloor \frac{n}{2^{f+1}} \rfloor \leq N(R_f) \leq    \lceil \frac{n}{2^{f+1}} \rceil.
\end{align}
%\textit{Example~4:}
%For a sequence of length three, the following conditions should be satisfied 
%\begin{align}
% \lfloor \frac{3}{4} \rfloor \leq N(R_1) \leq    \lceil \frac{3}{4} \rceil
%\end{align}
%\begin{align}
 %\lfloor \frac{3}{8} \rfloor \leq N(R_2) \leq    \lceil \frac{3}{8} \rceil.
%\end{align}
%The $x=(-1,-1,+1)$ is an m-sequence and have a run of length one and  a run of length two, hence, satisfies above conditions.

The ideal autocorrelation property and the run property are known to be independent for more than few decades until in 2009 Cai \cite{Auto-run} by thinking about autocorrelation run by run instead of symbol by symbol proved that these two  properties are related. The main result of his work can be presented in this relation \cite{Auto-run} 
 \begin{align}
  R_x(\tau)=n-2\tau\gamma-4\sum\limits_{f_1+f_2+\ldots+f_l<\tau}(-1)^l(\tau-i)N(R_{f_1}R_{f_2}\ldots R_{f_l})
 \end{align}
where $i=f_1+f_2+\ldots+f_l$, $\gamma$ is the total number of runs and $R_{f_1},R_{f_2},\ldots,R_{f_s}$ represent consecutive runs of lengths $f_1,f_2,\ldots,f_s$ in $x$.

Two special cases that can be obtained easily and would give us some understanding of run structure are $R_x(1)=n-2\gamma$ and $R_x(2)=n-4\gamma+4N(R_1)$. Therefore, sequences with ideal autocorrelation property have $\frac{n+1}{2}$ number of runs in which $\frac{n+1}{4}$ number of them are of length one. With this in mind, it may be true that the only sequences with ideal autocorrelation property that satisfy (\ref{38}) are m-sequences (Golomb's conjecture about m-sequences) but all the sequences that have ideal autocorrelation property are not too far from satisfying the conditions in (\ref{38}).

\textit{Example~4:}

If $n=11$ then (\ref{38}) implies that $ 1  \leq N(R_1) \leq  3$, $ 1  \leq N(R_2) \leq  2$, and $ 0  \leq N(R_f) \leq 1 $ for $f=3,\ldots,10$.
The codeword $x=(-1,-1,-1,~1,-1,-1,~1,-1,~1,~1,~1)$, which has the ideal autocorrelation property follows  (38) in all cases except $f=3$ (this codeword has two run of length three).

\begin{table}
	\caption{PN-sequences with ideal autocorrelation property of length less than 31 }
	\centering
	\begin{tabular}{*{7}{|c}|}
		\hline
		n& Sequence  & Type \\ \hline\hline
		3& (~1,~1,-1)
		& m-sequence  \\ \hline             
		
		7& (-1,-1,-1,1,-1,~1,~1)
		& m-sequence \\ & (-1,-1,-1,~1,~1,-1,~1)&  m-sequence \\ \hline
		
		11& (-1,-1,-1,~1,-1,-1,~1,-1,~1,~1,~1)  & Legendre\\ & (-1,-1,-1,~1   -1,~1,~1,-1,~1,~1,~1) & Legendre\\ \hline
		
		15& (-1,-1,-1,-1,~1,~1,~1,-1,~1,~1,-1,-1,~1,-1,~1)  & m-sequence\\ & (-1,-1,-1,-1,~1,-1,~1,-1,-1,~1,~1,-1,~1,~1,~1) & m-sequence\\ \hline
		19& (-1,-1,-1,-1,~1,-1,~1,-1,~1,~1,~1,~1,-1,-1,~1,-1,-1,~1,~1)  & Legendre\\ & (-1,-1,-1,-1,~1,-1,~1,-1,~1,~1,~1,~1,-1,-1,~1,~1,-1,~1,~1)   & Legendre\\ \hline
		23& (1,1,1,1,-1,1,-1,1,1,-1,-1,1,1,-1,-1,1,-1,1,-1,-1,-1,-1,-1)  & Legendre\\ & (1,-1,1,-1,-1,1,1,-1,-1,1,1,-1,1,-1,1,1,1,1,-1,-1,-1,-1,-1)   & Legendre\\ \hline

	\end{tabular}
\end{table}

\section{Conclusion}
 We investigated PN-sequences with ideal autocorrelation property and the consequence of this property on the number of $+1$s and $-1$s  and run structure of sequences. A new perspective was introduced using circulant matrix representation of PN-sequences. We derived a system of non-linear equations which led to ideal autocorrelation property from this point of view. Rewriting PN-sequence and its autocorrelation property in $\{0, 1\}$ led in a definition based on Hamming weight and Hamming distance and easily proved a number of results on PN-sequences with ideal autocorrelation property.

\end{document}